\documentclass[aps,prl,reprint,superscriptaddress,showpacs]{revtex4-1}

\usepackage{color}
\usepackage{graphicx}
\begin{document}

\title{H$_4$O and other hydrogen-oxygen compounds at giant-planet core pressures}

\author{Shuai Zhang}
\email[]{shuai.zhang01@berkeley.edu}
\author{Hugh F. Wilson}
\author{Kevin P. Driver}
\affiliation{Department of Earth and Planetary Science, University of California, Berkeley, California 94720, USA}
\author{Burkhard Militzer}
\affiliation{Department of Earth and Planetary Science, University of California, Berkeley, California 94720, USA}
\affiliation{Department of Astronomy, University of California, Berkeley, California 94720, USA}


\begin{abstract}
Water and hydrogen at high pressure make up a substantial fraction of the interiors of giant planets. Using {\it ab initio} random structure search methods we investigate the ground-state crystal structures of water, hydrogen, and hydrogen-oxygen compounds. We find that, at pressures beyond 14 Mbar, excess hydrogen is incorporated into the ice phase to form a novel structure with H$_4$O stoichiometry. We also predict two new ground state structures, $P2_1/m$ and $I4/mmm$, for post-$C2/m$ water ice.
\end{abstract}


\maketitle

\section{}
As one of the most abundant substances in the solar system, water ice at high pressure \cite{Polian1984, Benoit1996} is of fundamental importance in planetary science \cite{hubbard1984}. Over the last two years, a significant effort has been devoted to finding the ground-state structures of ice at the multimegabar pressures corresponding to the cores of gas giant planets (15-20 Mbar for Saturn and 40-60 Mbar for Jupiter). Militzer and Wilson \cite{Militzer2010} demonstrated that the $Pbcm$ phase \cite{Benoit1996} becomes unstable above 7.6 Mbar and predicted a new structure of $Pbca$ symmetry. Wang {\it et al}. \cite{Wang2011} showed that, at 8.1 Mbar, water ice assumes another structure of $I\bar{4}2d$ symmetry instead of transforming into a $Cmcm$ structure as proposed in Ref. \cite{Militzer2010}. Refs. \cite{McMahon2011,Ji2011,Hermann2011} instead proposed a structure with $Pmc2_1$ symmetry to appear in the same pressure regime, but it has a higher enthalpy than the $I\bar{4}2d$ structure. Around 12 Mbar, a transition to a structure with $P2_1$ symmetry was consistently predicted in Refs. \cite{Wang2011,McMahon2011,Ji2011,Hermann2011}. Even higher pressure studies predict a  $P2_1/c$ structure at 20 Mbar \cite{Ji2011}, and a metallic $C2/m$ structure at 60 Mbar \cite{McMahon2011,Hermann2011}. In nature, however, high-pressure water phases are rarely found in isolation, and, in gas giants, an icy core may be surrounded by a vast reservoir of hydrogen-rich fluid or exist in a mixture with the other planetary ices, such as ammonia and methane. Recent works have emphasized the fact that counterintuitive stoichiometries can occur in post-perovskite materials at extreme pressures \cite{Umemoto2011}.  This raises the question of whether H$_2$O is indeed the ground-state stoichiometry for water at high pressure in hydrogen-rich environments, or whether pressure effects are likely to result in the incorporation of hydrogen into the ice lattice to form novel ice-like phases with non-H$_2$O stoichiometry.

In this Letter, we apply {\it ab initio} random structure searching (AIRSS) methods \cite{Pickard2011} to determine the ground-state structure of hydrogen-oxygen compounds at extreme pressure to explore whether the H$_2$O stoichiometry of water is maintained at high pressure. The AIRSS method relies on the generation of a large number of random geometries whose structures are then optimized using density functional theory. In the AIRSS process, randomly generated unit cells are filled with randomly positioned atoms, and the structures are geometrically relaxed to the targeted pressure. Structures with competitive enthalpy are picked out and re-evaluated with more accurate thermodynamic calculations, from which the most stable structure can be determined.  
Although the AIRSS is not guaranteed to find the most stable phase, it has achieved remarkable success in discovering structures across a wide range of materials \cite{Pickard2006,*Pickard2008,Morris2009,Fortes2009}.

In our calculations, structural optimization is performed using the Vienna {\it ab initio} simulation package ({\footnotesize VASP}) \cite{kresse96a,*kresse96b}. Projector augmented wave (PAW) pseudopotentials \cite{kresse99} and the Perdew-Burke-Ernzerhof \cite{Perdew96} exchange-correlation functional are used. The pseudopotential cutoff radii equal to 0.8 and 1.1 Bohr for H and O. 
For efficiency of the AIRSS, a cutoff energy of 900 eV is chosen for initial relaxation, which is then increased to 1700 eV for precise calculations on favorable structures. Likewise, a relatively sparse Monkhorst-Pack grid \cite{monkhorst76,pack77} ($6\times6\times6$) is used initially, which is later replaced by a $20\times20\times20$ grid. Tests at the largest energy cutoff and grid size show that the total system energies are converged to the order of 1 meV, while the energy difference is converged to 0.1 meV.

 \begin{figure*}
 \includegraphics[width=1.0\textwidth]{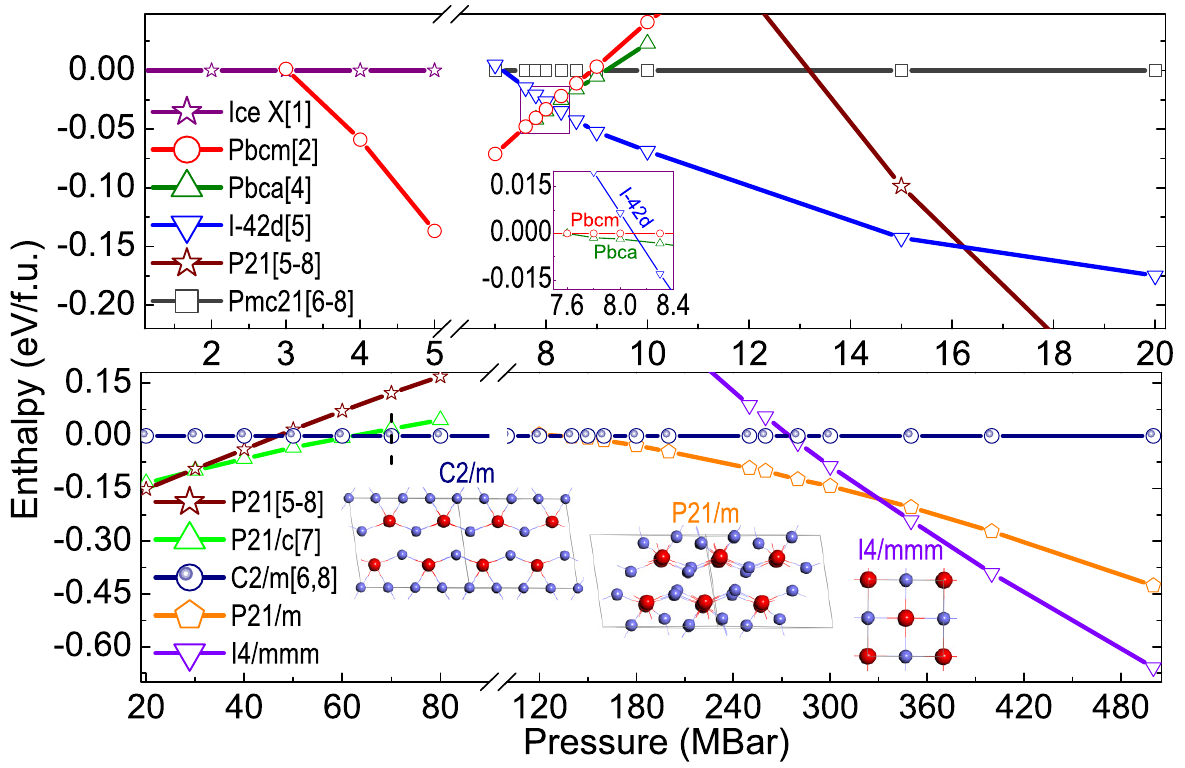}%
 \caption{\label{h2o-enth}(color online) Enthalpy versus pressure of different water ice structures. The reference following the name of each phase is the corresponding earliest report. In the lower panel,  structures are provided for $C2/m$, $P2_1/m$ and $I4/mmm$, where large red and small blue spheres represent O and H atoms, respectively. The dashed line at 70 Mbar indicates the highest pressure explored in previous work.}
 \end{figure*}

As an initial step, we recomputed enthalpies of recently reported \cite{Militzer2010,McMahon2011,Ji2011,Wang2011} H$_2$O structures as a function of pressure, assembling all proposed structures for comparison. Figure \ref{h2o-enth} shows the transition sequence up to the highest reported pressure (70 Mbar) as follows: $X\rightarrow Pbcm \rightarrow Pbca \rightarrow I\bar{4}2d \rightarrow P2_1 \rightarrow P2_1/c \rightarrow C2/m$, consistent with previous reports. Note that the $Pmc2_1$ structure predicted in Refs. \cite{McMahon2011,Ji2011,Hermann2011} is less stable than the $I\bar{4}2d$ structure reported by Wang {\it et al}. \cite{Wang2011} at all pressures; as the $I\bar{4}2d$ structure has eight formula units (f.u.) per unit cell, it was less likely to be generated by random search algorithms.

With quickly increasing number of known massive exoplanets, it is worthwhile to explore stable H$_2$O structures at even higher pressure. We use AIRSS to generate between 1000 and 3500 H$_2$O unit cells with 1-4 f.u. at pressures of 100, 150, 300, 400, and 500 Mbar. Additionally, we generate between 750 and 1150 structures with 5-8 f.u. at pressures of 100 and 200 Mbar. 

Figure \ref{h2o-enth} shows two new, stable phases with $P2_1/m$ and $I4/mmm$ symmetry that appear at high pressure. More stable than $C2/m$ above 135 Mbar, the $P2_1/m$ phase is also a monoclinic, layered structure that resembles the $C2/m$ phase except that O-H-O bonds are divided into two classes: one is shortened and straight, while the other is squeezed and distorted in different directions. At 330 Mbar, $P2_1/m$ is replaced by a tetragonal $I4/mmm$ phase, which has the same structure as the L'2 phase of ThH$_2$ \cite{ThH2}. $I4/mmm$ remains the most stable structure up to 500 Mbar. Like the $C2/m$ structure, an analysis of the electronic band structure shows that the $P2_1/m$ and $I4/mmm$ phases are metallic. Structural parameters of these two phases are listed in Supplemental Table I.

\begin{figure}
 \includegraphics[width=.5\textwidth]{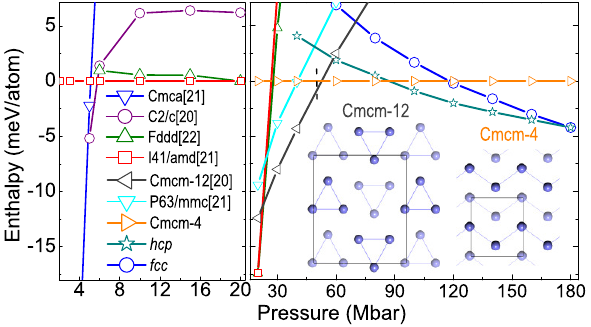}%
\caption{\label{h-enth}(color online) Enthalpy versus pressure of different structures of hydrogen relative to $I4_1/amd$ \cite{McMahon2010} in the left panel and the 4-atom $Cmcm$ phase on the right. The reference for each phase is given in the legend. In the right panel, structures of the 12-atom $Cmcm$ of Ref. \cite{Liu2012} and our new 4-atom $Cmcm$ are compared. The dashed line at 50 Mbar indicates the highest pressure explored in previous work.}
 \end{figure}

In addition to our initial studies on water ice, we performed AIRSS on hydrogen (results see Fig. \ref{h-enth}). Below 50 Mbar we find the same progression of ground state structures, $Cmca \rightarrow C2/c \rightarrow I4_1/amd \ {\rm (or} \ Fddd{\rm)} \rightarrow Cmcm$,  as has been reported previously \cite{Liu2012,McMahon2010,Geng2012}. When pressure exceeds 54 Mbar, we find that the 12-atom per unit cell $Cmcm$ structure of Ref. \cite{Liu2012}, hereafter $Cmcm$-12, becomes less stable than a new $Cmcm$-4 structure, in which three groups of bonds (different in length) exist without forming in-plane H$_3$ clusters \cite{Liu2012}. As pressure increases to 87 Mbar, the $hcp$ structure (symmetry $P6_3/mmc$) becomes stable. Different from the $P6_3/mmc$ structure predicted in Ref. \cite{McMahon2010} ($c/a\approx2$), our $hcp$ lattice exhibits a smaller $c/a$ ratio of 1.58. The $hcp$ structure remains stable until being replaced by the $fcc$ lattice at 180 Mbar.

Having computed the ground state structures of H and H$_2$O throughout a wide pressure range, we now explore novel hydrogen-oxygen stoichiometries, including hydrogen-enriched structures H$_3$O, H$_4$O, H$_6$O and H$_8$O and oxygen-rich structures HO, HO$_2$ and HO$_4$. Between 200 and 3000 random structures with up to 8 f.u. at pressures of 20-500 Mbar were generated, depending on the specific computational cost and searching efficiency. 

\begin{figure}
\includegraphics[width=.5\textwidth]{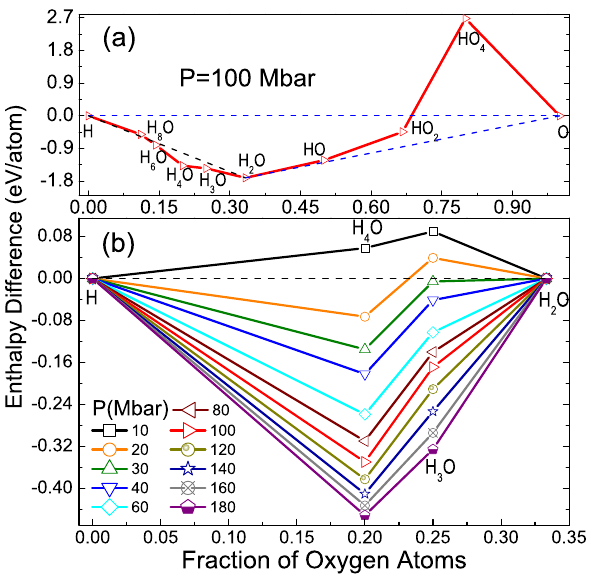}%
 \caption{\label{ho-CH}(color online) (a) Enthalpy of different H-O compounds at 100 Mbar, relative to a mixture of pure H and O. (b) Enthalpy of hydrogen-rich H-O compounds at several pressures, relative to a mixture of H and H$_2$O.}
 \end{figure}

Figure \ref{ho-CH} compares the stability of structures with different stoichiometries using a convex hull diagram \cite{Morris2011}. This diagram shows the per-atom enthalpy of the compound H$_m$O$_n$, as a function of fractional concentration of oxygen, $x=n/(m+n)$, relative to a combination of the two phases at the ends. A particular H$_m$O$_n$ phase is stable relative to a mixture of two phases of different concentrations if its enthalpy lies below the line that interpolates between those two phases. In Fig. \ref{ho-CH}(a) we plot a convex hull diagram showing the enthalpies of possible H-O phases at a pressure of 100 Mbar. The H$_4$O structure emerges as the most stable on the hydrogen-enriched side of the diagram, implying that any H-O complex with an H:O concentration greater than 2:1 will preferentially form an H$_4$O+H or an H$_4$O+H$_2$O mixture, in contrast to the H$_2$O+H$_2$ mixture we expect at ambient pressure. On the oxygen-rich side of the graph, the HO, HO$_2$ and HO$_4$ phases are unstable relative to H$_2$O+O mixtures. 

\begin{figure}
 \includegraphics[width=0.5\textwidth]{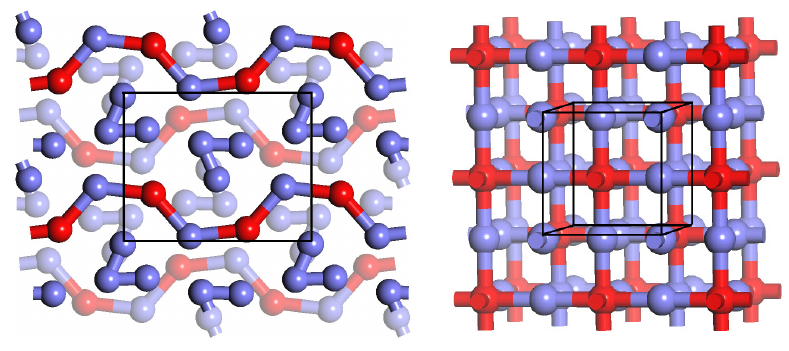}%
 \caption{\label{ho-struc}(color online) Structures of H$_4$O-$Pnma$ (left) and H$_3$O-$bcc$ (right). Red and blue spheres denote O and H atoms, respectively.}
 \end{figure}

In Fig. \ref{ho-CH}(b) we focus on convex hull for the hydrogen-enriched compounds including the H$_3$O and H$_4$O phases after having switched the right end from O to H$_2$O. Besed on our calculations at 10 and 20 Mbar, we predict H$_4$O to become stable relative to a mixture of H and H$_2$O at 14 Mbar. Such a pressure is predicted to for the core-mantle boundary (CMB) of Saturn and far below that of Jupiter. For pressures above 140 Mbar, we find H$_3$O becoming stable relative to H$_2$O+H$_4$O mixtures. These results imply that at giant-planet CMB pressures, in the absence of temperature effects, excess hydrogen present is incorporated into existing water ice structures to form the novel hydrogen-rich compound, H$_4$O. On the other hand, no oxygen-rich compounds will form, which is in agreement with recent calculations at high temperatures \cite{Wilson2012}.

The H$_4$O phase that we find here is a layered orthorhombic structure with $Pnma$ symmetry. Each layer is characterized by -H-O-H-O- chains, of which each O is associated with one H$_3$ unit shown in Fig. \ref{ho-struc}. Our band structure analysis showed that the H$_4$O structure becomes metallic at 80 Mbar. For H$_3$O, we find a $bcc$ ($Im\bar{3}m$) structure above 110 Mbar, a layered monoclinic $C2/m$ structure (see Supplemental Fig. 1) at  20-110 Mbar, and a slightly less stable orthorhombic $Cmmm$ structure (see Supplemental Fig. 2) between 20 and 100 Mbar, whose enthalpy is very close (2-36 meV/f.u.) to that of the $C2/m$ structure.

To verify the applicability of the PAW pseudopotential under extreme pressures, we performed all-electron LAPW calculations using the {\footnotesize ELK} code \cite{elk}. The LAPW calculations for hydrogen at 80-180 Mbar reconfirm the $Cmcm \rightarrow hcp$ transition at 87 Mbar and increase the enthalpy of the $fcc$ structure by 5 meV, precluding $fcc$ to become stable until a pressure of about 200 Mbar is reached. We also recalculated the relative enthalpy of two systems (H+H$_2$O versus H$_3$O, at 100 Mbar), and found that the difference between the two methods is only 4 meV/f.u.. We further replaced the {\footnotesize VASP}-PAW pseudopotential of oxygen with an all-electron PAW potential (cutoff radius equals to 0.95 Bohr), re-optimized two H$_2$O structures, $I4/mmm$ and $C2/m$, and compared their enthalpy. The enthalpy difference, between the $I4/mmm$ and the $C2/m$ structures, increases by 16 and 70 meV/f.u. at 180 and 400 Mbar, respectively. All these corrections are small and confirm our prediction of two new ice phases, although their transition pressures increase slightly (by $< 4\%$ for $P2_1/m$, $< 18\%$ for $I4/mmm$).

We also estimated the influence of zero-point motion (ZPM) by phonon calculations with $2\times2\times2$ supercells using the finite displacement method \cite{phon}. Including zero-point energy (see Supplemental Table II), the convex hull diagram in Fig. \ref{ho-CH}(b) shifts in the positive direction by 12 meV/aotm for H$_4$O and 3 meV/aotm for H$_3$O at 20 Mbar, and by 74 meV/aotm for H$_4$O and 89 meV/aotm for H$_3$O  at 180 Mbar. By applying these changes to the linearly interpolated enthalpy, we estimate the ZPM increases the pressure at which H$_4$O and H$_3$O become more stable than their corresponding H+H$_2$O mixtures by, at most, 1.3 Mbar. Relative to the H$_2$O+H$_4$O mixture, H$_3$O becomes less stable below 180 Mbar. We note that, based on the findings in Refs. \cite{McMahon2011, Hermann2011}, the ZPM-induced changes in the enthalpy differences between H$_2$O structures are small and do not change the sequence of phase transition, although the transition pressures will be slightly affected. Further, the difference in the zero-point energy of different hydrogen structures is small ($<80$ meV/atom when below 50 Mbar), as indicated in Ref. \cite{McMahon2010}, and will not change our conclusions on the stability of H$_4$O.

In this work we have replicated and extended the phase diagram of solid hydrogen to 180 Mbar and that of water ice to 500 Mbar. We also predict novel, hydrogen-enriched meta-ice structures to become more stable than hydrogen-water mixtures at high pressure. Our results imply that water ice under conditions in excess of 14 Mbar at low temperature will absorb excess hydrogen from its environment to form a hydrogen-rich H$_4$O phase. The 14 Mbar pressure in question is comparable to the CMB of Saturn and far below the core pressure of Jupiter, and may potentially also be reached inside super-Neptune ice giants. This result stands in contrast to the tendency of methane to preferentially expel hydrogen from its own structure under pressure to form hydrocarbons and eventually diamond \cite{Ancilotto1997}, and in ice mixtures methane could potentially provide the excess hydrogen to incorporate into H$_4$O. These results underline the fact that chemistry at high pressure may deliver substantially counterintuitive results, and that consideration of structures likely to form at giant-planet core conditions requires looking beyond traditional ambient-pressure chemistry to explore unfamiliar stoichiometries and combinations of elements.


%

\end{document}